# Effect of antifluorite layer on the magnetic order in Eu-based 1111 compounds, Eu$T$AsF ($T$ = Zn, Mn, and Fe)

Igor V. Plokhikh,[1*] Alexander A. Tsirlin,[2] Dmitry D. Khalyavin,[3] Henry E. Fischer,[4] Andrei V. Shevelkov[5] and Arno Pfitzner[6]

[1]Laboratory for Multiscale Materials Experiments, Paul Scherrer Institut, PSI, Villigen, CH-5232, Switzerland

[2]Felix Bloch Institute for Solid-State Physics, University of Leipzig, 04103 Leipzig, Germany

[3]ISIS Facility, Rutherford Appleton Laboratory, Harwell Oxford, Didcot OX11 0QX, UK

[4]Institut Laue-Langevin, 71 avenue des Martyrs, CS 20156, 38042 Grenoble Cédex 9, France

[5]Department of Chemistry, Lomonosov Moscow State University, 119991 Moscow, Russia

[6]Institute of Inorganic Chemistry, University of Regensburg, 93053 Regensburg, Germany

*Corresponding author: igor.plohih@gmail.com, igor.plokhikh@psi.ch

**Abstract**

The 1111 compounds with an alternating sequence of fluorite and antifluorite layers serve as structural hosts for the vast family of Fe-based superconductors. Here, we use neutron powder diffraction and density-functional-theory (DFT) band-structure calculations to study magnetic order of $Eu^{2+}$ in the $[EuF]^+$ fluorite layers depending on the nature of the $[TAs]^-$ antifluorite layer that can be non-magnetic semiconducting ($T$ = Zn), magnetic semiconducting ($T$ = Mn), or magnetic metallic ($T$ = Fe). Antiferromagnetic transitions at $T_N \sim 2.4 - 3$ K due to an ordering of the $Eu^{2+}$ magnetic moments were confirmed in all three Eu$T$AsF compounds. Whereas in Eu$T$AsF ($T$ = Zn and Mn), the commensurate $k_1$ = (½ ½ 0) stripe order pattern with magnetic moments within the $a$-$b$ plane is observed, the order in EuFeAsF is incommensurate with $k$ = (0 0.961(1) ½) and represents a cycloid of $Eu^{2+}$ magnetic moments confined within the $bc$-plane. Additionally, the $Mn^{2+}$ sublattice in EuMnAsF features a robust $G$-type antiferromagnetic order that persists at least up to room temperature, with magnetic moments along the $c$-direction. Although DFT calculations suggest stripe antiferromagnetic order in the Fe-sublattice of EuFeAsF as the ground state, neutron diffraction reveals no evidence of long-range magnetic order associated with Fe. We show that the frustrating interplane interaction $J_3$ between the adjacent $[EuF]^+$ layers is comparable with in-plane $J_1$-$J_2$ interactions already in the case of semiconducting fluorite layers $[TAs]^-$ ($T$ = Zn and Mn) and becomes dominant in the case of the metallic $[FeAs]^-$ ones. The latter, along with a slight orthorhombic distortion, is proposed to be the origin of the incommensurate magnetic structure observed in EuFeAsF.

**Introduction**

Interplay between magnetism and superconductivity in the family of iron-based superconductors is of particular interest as iron itself and many of its compounds are ferromagnets, *i.e.* they exhibit properties that are detrimental for superconductivity.[1,2] The general structural feature of iron-based superconductors is the presence of negatively charged antifluorite layers $[FeAs]^-$ (2D infinite motif of edge-sharing tetrahedra $FeAs_4$) that are separated by positively charged counter-layers.[3–5] In the majority of compounds, these counter-layers act only as charge reservoirs and affect the behaviour of the functional FeAs-layers only in an indirect manner, *i.e.* by adjusting geometry or charge carriers sign and their concentration. They are usually ionic and hence do not contribute to the electronic structure at the vicinity of the Fermi level. A wide choice of positively charged layers leads to a crystallographic diversity of iron-based superconductors – more than ten structure types have been reported until now.[6]

Examples of FeAs-based superconductors containing magnetically active ions in the counter-layer are sparse, but the existing cases trigger vivid discussions in the literature. The simplest possibility arises from placing magnetically active $Eu^{2+}$ ($4f^7$) cations solely between the FeAs layers, resulting in the 122-type compound, $EuFe_2As_2$.[7] Electron/hole doping through a substitution in the Eu and Fe positions, or chemical pressure via P-to-As substitution induce superconductivity with $T_C$ up to 29 K, *i.e.* the compound behaves as a common member of this family without any obvious influence of the $Eu^{2+}$ magnetism.[8] The most interesting difference, however, is that detwinning of magnetic domains below tetragonal-to-orthorhombic distortion temperature can be realized without mechanical straining, but through an application of moderate magnetic fields. This effect was ascribed to a biquadratic exchange coupling between the Fe and Eu spins.[9,10] In addition, several variants of doping stabilize an exotic ground state with coexisting superconductivity and ferromagnetism,[11] as also observed in $RbEuFe_4As_4$ that features a decorated variant of the 122 structure.[12,13]

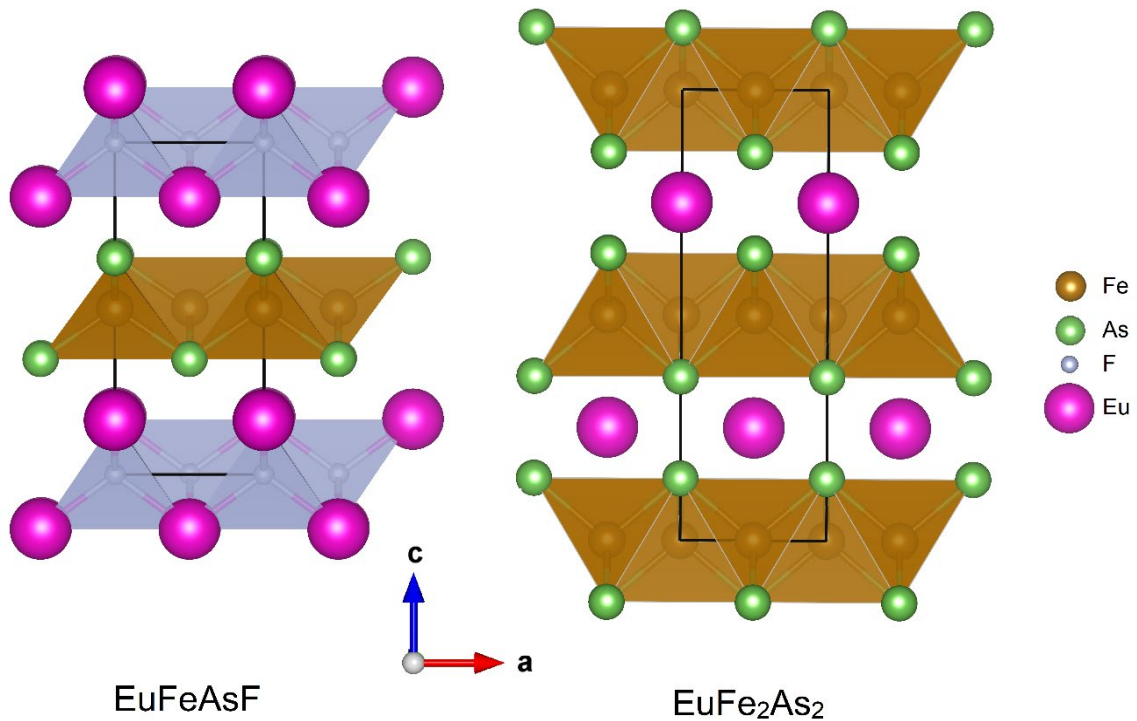

***Figure 1.*** Comparison of crystal structures of EuFeAsF (left) and $EuFe_2As_2$ (right).

Replacing the single tetragonal net of $Eu^{2+}$ cations with the fluorite layer $[EuF]^+$ leads to another compound, EuFeAsF, the representative of the 1111 (also known as LaOAgS or ZrCuSiAs) structure type.[14–16] A comparison of the 122 and 1111 structures is provided in ***Figure 1***. In the current work, we study the interplay between the fluorite-type $[EuF]^+$ layer with the high magnetic moment ($Eu^{2+}$, $4f^7$) and different types of antifluorite layers: non-magnetic semiconducting $[ZnAs]^-$ ($Zn^{2+}$, $3d^{10}$),[17] magnetic semiconducting $[MnAs]^-$ ($Mn^{2+}$, $3d^5$),[18] and magnetic metallic $[FeAs]^-$ ($Fe^{2+}$, $3d^6$) layers. This comparison allows us to identify dominant magnetic interactions in the Eu-based 1111 compounds. On general grounds, one may expect that $Eu^{2+}$ magnetism should be described by the $J_1$-$J_2$ Heisenberg model on the square lattice. Interestingly, we have found that this description holds indeed, but with $J_1$ replaced by $J_3$, the long-range interactions that goes across the antifluorite layers and shows strong variability across the 1111 family depending on the transition metal contained therein. Consequently, different types of the $Eu^{2+}$ magnetic order are realized.

Although neutron scattering is the most natural way to elucidate the magnetic ordering on a microscopic level and to follow its evolution with external parameters, such studies in the case of europium are hampered due to the strong neutron absorption of the natural isotopic mixture containing *ca.* half of $^{151}Eu$ with 9100 barn absorption cross-section for neutrons around the thermal region.[19] Still, as the effectiveness of modern neutron diffractometers (WISH at ISIS) grows, it becomes possible to acquire reasonable data within reasonable time using standard diffractometer setups, which is also facilitated by the high magnetic moment of $Eu^{2+}$. Another well-known approach, also used in this work, is to utilize hot neutrons (D4 at ILL), for which Eu absorption becomes bearable.[20]

**Experimental part**

*Synthesis and primary characterization.* Large powder samples of EuZnAsF, EuMnAsF, and EuFeAsF were produced through a standard ampoule technique, similar to that described in ref. 17 and 18. Elemental Eu (Chempur, 99.5%), $EuF_3$ (Chempur, 99.999%), Zn (Chempur, 99.9%), Mn (Chempur, 99.9%), As (Chempur, 99.9%) and Fe (Alfa Aesar, 99.999%) were used as starting materials. Stoichiometric mixtures were annealed at 1173 K in evacuated silica tubes several times until homogenous black powders were obtained. This synthetic procedure leads to nearly phase pure samples without using binary precursors as it has been described in.[14,15] The phase composition was checked using a STOE STADI P diffractometer (STOE & Cie, Darmstadt, Germany) equipped with a Dectris Mythen 1K detector with Cu-K$\alpha_1$ radiation ($\lambda$ = 1.5406 Å).

*Neutron diffraction.* Neutron powder diffraction data were collected at the ISIS pulsed neutron and muon facility of the Rutherford Appleton Laboratory (UK) on the WISH time-of-flight long-wavelength diffractometer located at the second target station (TS2)[21] as well as at the Institut Laue-Langevin (research nuclear reactor) on the D4 hot-neutron powder diffractometer.[22] For the WISH experiment, the samples of EuZnAsF, EuMnAsF, and EuFeAsF (~1 g each) were loaded into cylindrical 3mm V-cans and measured in the temperature range of 1.5 – 300 K. For the D4 experiment, the sample of EuFeAsF (~3.5 g) was loaded in a 5mm V-can and measured in the temperature range of 2.4 – 300 K using 0.7 Å neutrons. The raw data from both experiments are available under refs. 23 and 24. Crystal and magnetic structures solution and refinement were performed using the JANA2006 [25,26] and FullProf[27] software using magnetic space group-subgroup relations and representation analysis. Absorption correction for WISH data was done using empirical model implemented in JANA2006, whereas D4 data were corrected explicitly assuming cylindrical sample and measured transmission coefficient of 9%. Double-$k$ magnetic models for EuMnAsF were constructed using ISODISTORT.[28,29] Monte Carlo simulations of magnetic diffuse scattering has been done using the Spinvert code.[30]

*Microscopic modelling.* Exchange couplings were calculated by a mapping procedure using total energies obtained from density-functional calculations performed in the FPLO code[31] with the Perdew-Wang exchange-correlation potential.[32] In the case of Eu $4f$ states, the mean-field DFT+$U$ correction with double-counting in the fully localized limit was applied. Thermodynamic properties were obtained from classical Monte-Carlo simulations performed using the *spinmc* algorithm of the ALPS package.[33]

**Results and discussion**

Magnetic and transport properties as well as heat capacity data for EuZnAsF and EuMnAsF were reported previously, [17,18] whereas magnetization data for EuFeAsF were measured in the current study and are provided in SI. With respect to the magnetization data, all compounds are qualitatively similar; they are paramagnetic down to helium temperatures with the magnetic moment of 7.94 $\mu_B$ derived from the Curie-Weiss fits and consistent with the effective moment expected for pure $Eu^{2+}$. Sharp peaks are observed at $T_N$ = 2.6 K in EuZnAsF, $T_N$ = 3.0 K in EuMnAsF, and $T_N$ = 2.9 K in EuFeAsF. These peaks manifest antiferromagnetic ordering in the $Eu^{2+}$ sublattice as proved by neutron diffraction data (*vide infra*). Additionally, in all samples, magnetic susceptibility measurements in low fields (below 0.1T) reveal a further effect manifested by small kinks at 5-5.5 K. However, neutron diffraction data revealed no evidence of the transition around 5 K. Therefore, we identify these kinks as effects unrelated to the magnetic ordering. They may originate from either short-range correlations or impurities/defects.

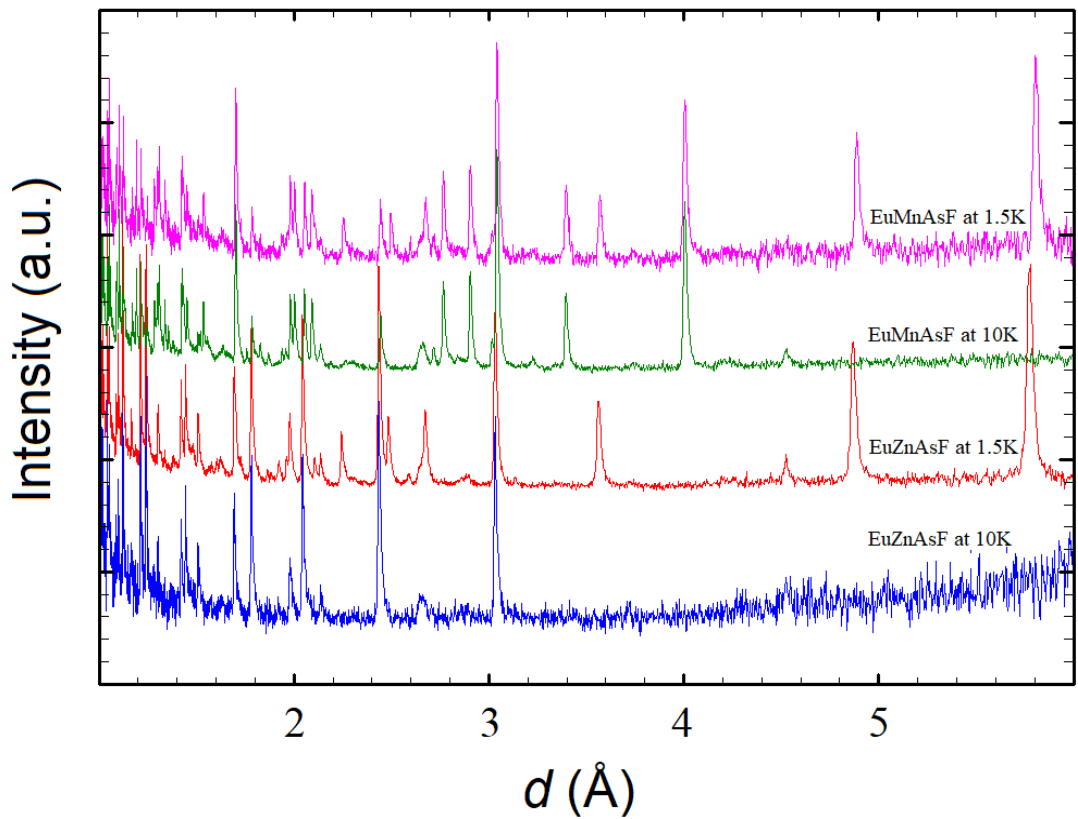

*Figure 2.* From bottom to top: Comparison of normalized time-of-flight (TOF) neutron diffraction patterns of EuZnAsF at 10 K in the paramagnetic state, EuZnAsF at 1.5 K with antiferromagnetic ordering in the $Eu^{2+}$ sublattice, EuMnAsF at 10 K with antiferromagnetic ordering in the $Mn^{2+}$ sublattice, and EuMnAsF at 1.5 K with antiferromagnetic ordering of both $Mn^{2+}$ and $Eu^{2+}$. Note that diffraction patterns can be directly compared to each other, owing to the similar unit-cell metrics of both compounds.

The trial neutron diffraction measurements show that the employed setups provide data sets of a quality sufficient for structure refinement, despite the strong neutron absorption. A comparison of the normalized diffraction patterns of Eu*T*AsF (*T* = Zn and Mn) at 1.5 K and at 10 K is provided in *Figure 2*. EuZnAsF at 10 K exhibits only nuclear reflections, but above 4 Å it exhibits an increased background, which is characteristic for paramagnetic compounds that approach a magnetic phase transition. Additional reflections appear upon cooling of this sample down to 1.5 K. They can be indexed using a commensurate propagation vector $k_1 = (½ ½ 0)$. EuMnAsF already above 4 K exhibits reflections that cannot be explained using the structural model consistent with the room-temperature X-ray data. They all can be indexed using the propagation vector $k_2 = (0\ 0\ ½)$. Further cooling of EuMnAsF down to 1.5 K leads to the development of additional reflections, which do not overlap with those of $k_2$, but lie very close to those described by $k_1$. The magnetic model in this case should be a superposition of the two, without any evident mutual influence.

The models of magnetic structures have been derived by testing all magnetic super space groups compatible with the crystallographic symmetry and the given propagation vectors; they are schematically shown in *Figure 3*. Magnetic moments of $Eu^{2+}$ are located at the vertices of tetrahedra of the fluorite structure. In EuZnAsF, the moments align with the *a*-direction of the magnetic unit cell ($\sqrt{2}a \times \sqrt{2}a \times c$ supercell of the crystallographic one). Two out of four $Eu^{2+}$ magnetic moments, located at the vertices of tetrahedra, are coupled ferromagnetically and together with magnetic moments from the neighboring layer form a ferromagnetic stripe, propagating along the *c*-axis. Magnetic unit cell contains two of those ferromagnetic stripes stacked to each other in an antiferromagnetic manner along the *b*-axis, yielding the global antiferromagnetic order. From the perspective of the $J_1$-$J_2$ model, this magnetic order should be driven by $J_2$ because the respective bonds always show antiparallel spins. By contrast, the $J_1$ bonds feature both parallel and antiparallel spins, indicating that $J_1$ does not contribute to stabilizing magnetic order.

The magnetic structure of EuMnAsF at 10 K closely resembles that of the previously reported BaMn*Pn*F as well as other Mn-containing 1111 materials, [34,35] where G-type magnetic ordering is observed. Magnetic moments of $Mn^{2+}$ cations centering the $MnAs_4$ tetrahedra are aligned along the *c*-direction. Each $Mn^{2+}$ magnetic moment is coupled antiferromagnetically to its nearest neighbors, forming a checkerboard layer. These layers are stacked antiferromagnetically along the *c*-direction. The refined magnetic moment at 10 K is 3.4 $\mu_B$. It is lower than the expected value of 5 $\mu_B$ for the isolated $Mn^{2+}$ ($3d^5$) cation, but in a good agreement with other 1111 compounds, where refined moments are in the range between 3 and 3.9 $\mu_B$.

In agreement with the full separation of the $k_1$ and $k_2$ magnetic reflections, magnetic structure of EuMnAsF at 1.5 K features independent $Eu^{2+}$ and $Mn^{2+}$ orders in the $\sqrt{2}a \times \sqrt{2}a \times 2c$ magnetic supercell according to the pattern described above. Also, several diffraction patterns measured between 4 K and RT show no changes of the propagation vector for the $Mn^{2+}$ sublattice. Together with the data obtained from thermodynamic measurements, this rules out a spin-reorientation scenario in the $Mn^{2+}$ sublattice, contrary to CeMnAsO and ThMnAsN.[36–40]

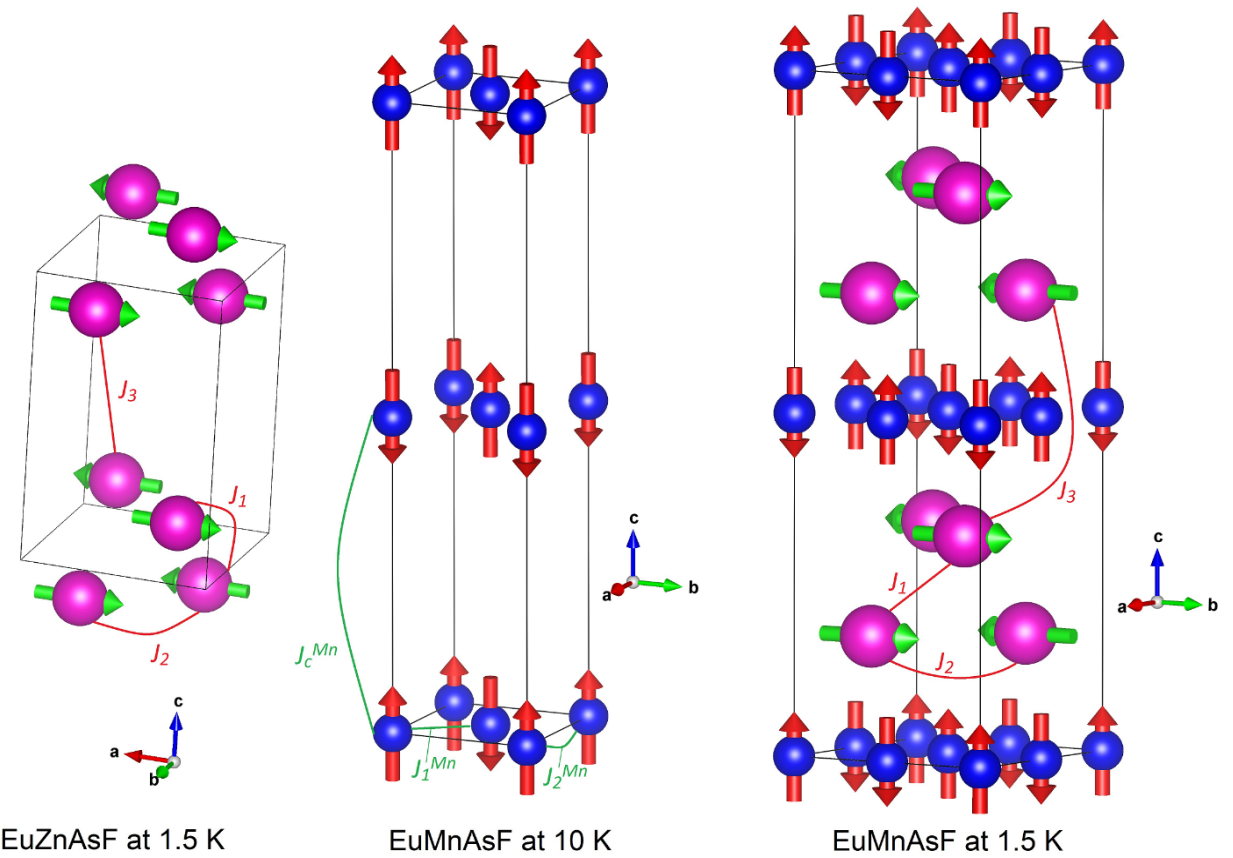

*Figure 3.* From left to right: Magnetic structure of EuZnAsF at 1.5 K, EuMnAsF at 10 K, and EuMnAsF at 1.5 K. Magnetic unit cells are outlined. Only magnetic atoms (Mn – blue, Eu – pink) with their ordered magnetic moments (red and green, respectively) are shown for clarity.

*Table 1.* Details of neutron powder diffraction experiments, magnetic structure refinements and a summary of magnetic properties measurements.

|  | EuZnAsF | EuMnAsF | EuMnAsF | EuFeAsF |
|---|---|---|---|---|
| Temperature | 1.5 K | 10 K | 1.5 K | 1.5 K |
| Nuclear space group | *P4/nmm* | *P4/nmm* | *P4/nmm* | *Cmme* |
| Cell parameters | | | | |
| $a$ in Å | 4.0852(2) | 4.1075(2) | 4.1059(2) | 5.5923(2) |
| $b$ in Å | = a | = a | = a | 5.6335(2) |
| $c$ in Å | 9.0462(3) | 9.0486(3) | 9.0447(5) | 8.8812(3) |
| k-vector | | | | |
| $k_1$ | (½ ½ 0) |  | (½ ½ 0) | (0 0.961(1) ½) |
| $k_2$ |  | (0 0 ½) | (0 0 ½) |  |
| Magnetic space group | $P_Cmab$[1] | $P_C4_2/ncm$ | $P_Ibcm$ | $Cmm2.1'(0b½)s00s$ |
| $d$-spacing range in Å | 0.75 – 5.85 | 0.8 – 4.8 | 0.96 – 6.1 | 1.15 – 5.70 |
| Reflections | 443/230[2] | 453/123[2] | 464/346[2] | 339/127[2] |
| Structural parameters | 7/1[2] | 6/1 | 9/2 | 7/1 |
| R values | | | | |
| $R_P$ in % | 3.19 | 3.10 | 5.52 | 3.84 |
| $R_{wP}$ in % | 2.83 | 3.42 | 5.80 | 4.72 |
| $\chi^2$ | 1.28 | 1.77 | 1.07 | 1.36 |
| $R_I$ in % | 4.41/3.90/6.23[3] | 2.07/1.63/3.23[3] | 5.47/4.88/7.17[3] | 5.55/4.55/7.32[3] |
| Refined magnetic moment in $\mu_B$ | 5.47(2)/$Eu^{2+}$ | 3.39(2)/$Mn^{2+}$ | 4.59(6)/$Eu^{2+}$ 3.56(5)/$Mn^{2+}$ | 5.64(3) $\mu_B$/$Eu^{2+}$ |
| $\mu_{CW}$ in $\mu_B$ | 7.93 | 7.67 | 7.67 | 7.92 |
| $M_S$ at 2K in $\mu_B$ | 7.00 | 6.84 | 6.84 | 6.75 |

[1] The non-standard setting is chosen to preserve the relation to initial paramagnetic phase
[2] Overall/magnetic
[3] For all reflections/ structural reflections/magnetic reflections

The diffraction patterns of EuFeAsF measured above (at 10 K) and below (at 1.5 K) the 3 K phase transition are shown in *Figure 4*. Several very strong magnetic reflections are observed, some of them are even stronger than the nuclear ones. The triplet of magnetic reflections at 5.19 Å, 5.33 Å, and 5.57 Å, at the *d*-spacings comparable though not coinciding with the *a* and *b* cell parameters, is indicative of an incommensurate magnetic order. Indeed, these reflections, as well as those at lower *d*-spacings, can be indexed with the single incommensurate vector $k$ = (0 0.961(1) ½) (assuming the following setting of the unit cell: $a$ = 5.5917(2) Å, $b$ = 5.6344(2) Å and $c$ = 8.8821(3) Å). Magnetic superspace and representation analysis reveal that the group $Cmm2.1'(0b^{1}/_{2})s00s$ having two symmetry independent Eu atoms in the unit cell, provides the best fit; description of alternative options is provided in SI. By forcing the same amplitude of the magnetic moments in both atomic positions, as well as the same amplitude throughout the whole structure, we obtain the final solution provided in *Table 1.* The refinement returned the magnetic moment of 5.64(3) $\mu_B$, which is only slightly lower than the expected value of 7 $\mu_B$. The deviation may be explained by the proximity of our measurement temperature to the Néel temperature, such that magnetic moments are not fully saturated. Similar to the aforementioned commensurate case, the *b*-components of the magnetic moments form two stripes per unit cell that are coupled to each other in antiferromagnetic manner. The magnetic moments gradually rotate in the *bc* plane when passing from one $Eu^{2+}$ ion to another along the *b*-direction. The overall spin arrangement in the *ab* plane resembles that in EuZnAsF (nearest-neighbor spins along *b* are antiparallel, *Figure 5*), albeit with an additional incommensurate modulation.

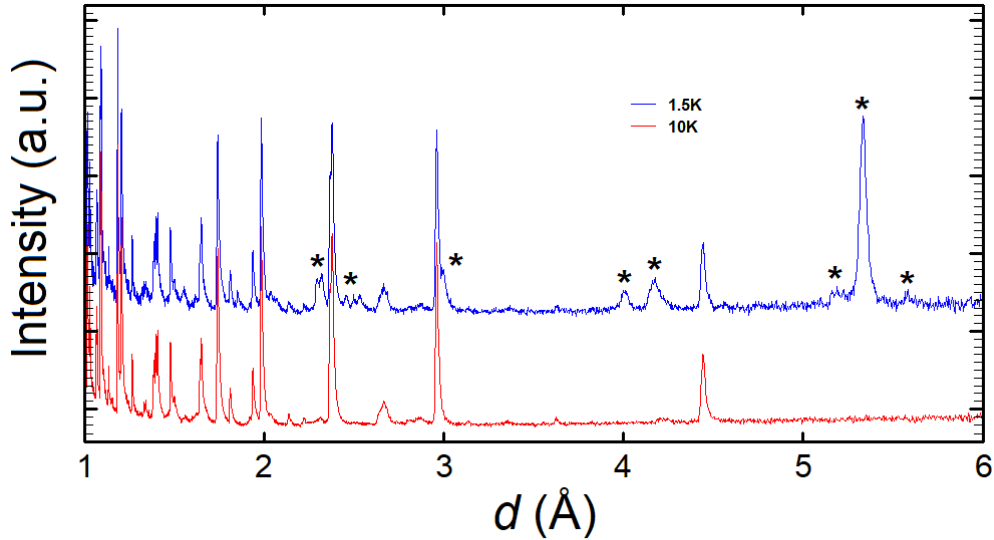

*Figure 4.* Comparison of TOF neutron diffraction patterns of EuFeAsF at 1.5 K (below magnetic ordering in the $Eu^{2+}$ sublattice) and at 10 K. The strongest magnetic reflections or their groups are outlined by asterisks.

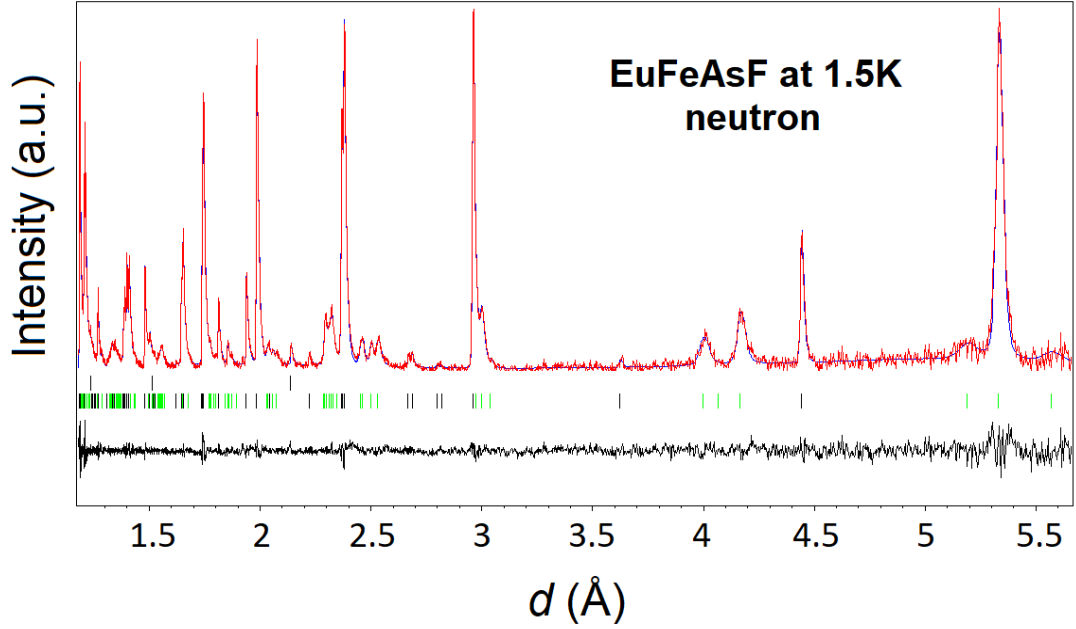

***Figure 5.*** Refinement of the nuclear and magnetic structures of EuFeAsF against the TOF neutron diffraction data at 1.5 K. The measured points are red, the calculated curve is blue, the positions of structural reflections and the difference curve are black, the positions of magnetic reflections are green. Reflections from the vanadium container were also included in the refinement.

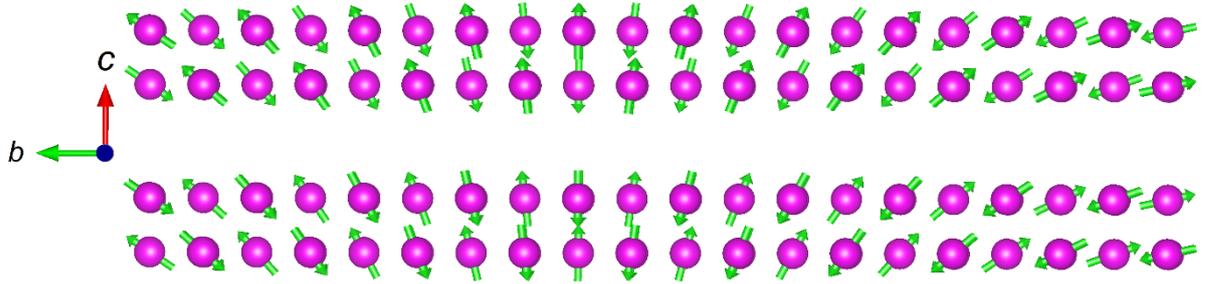

***Figure 6.*** $a \times 9b \times 2c$ supercell approximant for EuFeAsF. Only Eu atoms with their magnetic moments are shown for clarity.

Signatures of magnetic ordering in the Fe sublattice were observed neither in the WISH data, nor in the D4 measurement performed with very high statistics. In the sibling compound, SrFeAsF, the ordered moment is as low as 0.6 $\mu_B$,[41,42] which should be accessible using both instruments. Our data thus suggest that long-range order in the Fe sublattice is associated with a very low ordered moment or absent entirely, similar to what was observed in $EuFe_2P_2$.[43] Moreover, it is consistent with the absence of any obvious transition related to Fe spins in both magnetic susceptibility data provided in SI.

The presence of *dynamic* short-range spin-spin correlations in EuFeAsF already above the Néel temperature is evident from the pronounced magnetic diffuse scattering shown in ***Figure 7***. These plots are obtained assuming that all additional scattering in low-temperature patterns is due to the evolving magnetism. Although small oscillations are visible at least up to 30 K, the scattering seems to saturate when passing from the 10 K to the 6 K pattern, indicative of the formation of *static* short-range spin-spin correlations with some degree of ordering. This is in a good agreement with the Curie-Weiss temperature of -7 K that determines the energy scale of magnetic couplings. Further on, based on the diffuse scattering data we calculated the Eu-Eu spin correlation function

⟨**S**(0)·**S**(r)⟩, which is a vector product of magnetic moment of an atom at a given position with an atom at a given distance r, averaged over the whole structure. This function is equal to +1 (-1) when ferromagnetically (antiferromagnetically) coupled spins are fully co-aligned, whereas its absolute value can be reduced in real systems due to various effects.

The negative and positive values of ⟨**S**(0)·**S**(r)⟩ at, respectively, ca. 4 Å and 5.6 Å (***Figure 7***, right panel) are reminiscent of the $Eu^{2+}$ order in EuZnAsF and EuMnAsF where first neighbors in the *ab* plane (~4 Å, $J_2$) are antiparallel, and second neighbors in the *ab* plane (5.6 Å) are parallel by virtue of being connected via two $J_2$ bonds. The values at 5.6 Å are higher by absolute value because the spin-spin correlations at 4 Å also include the bonds $J_1$ where spins may be both parallel and antiparallel. Remarkably, the interlayer spin-spin correlations at 6.6-6.8 Å are quite weak. This illustrates that short-range order in EuFeAsF mostly involves antiferromagnetic order in the *ab* plane with the loss of coherence between the layers. Overall, spin arrangement above the transition temperature is consistent with the long-range-ordered magnetic structure described above.

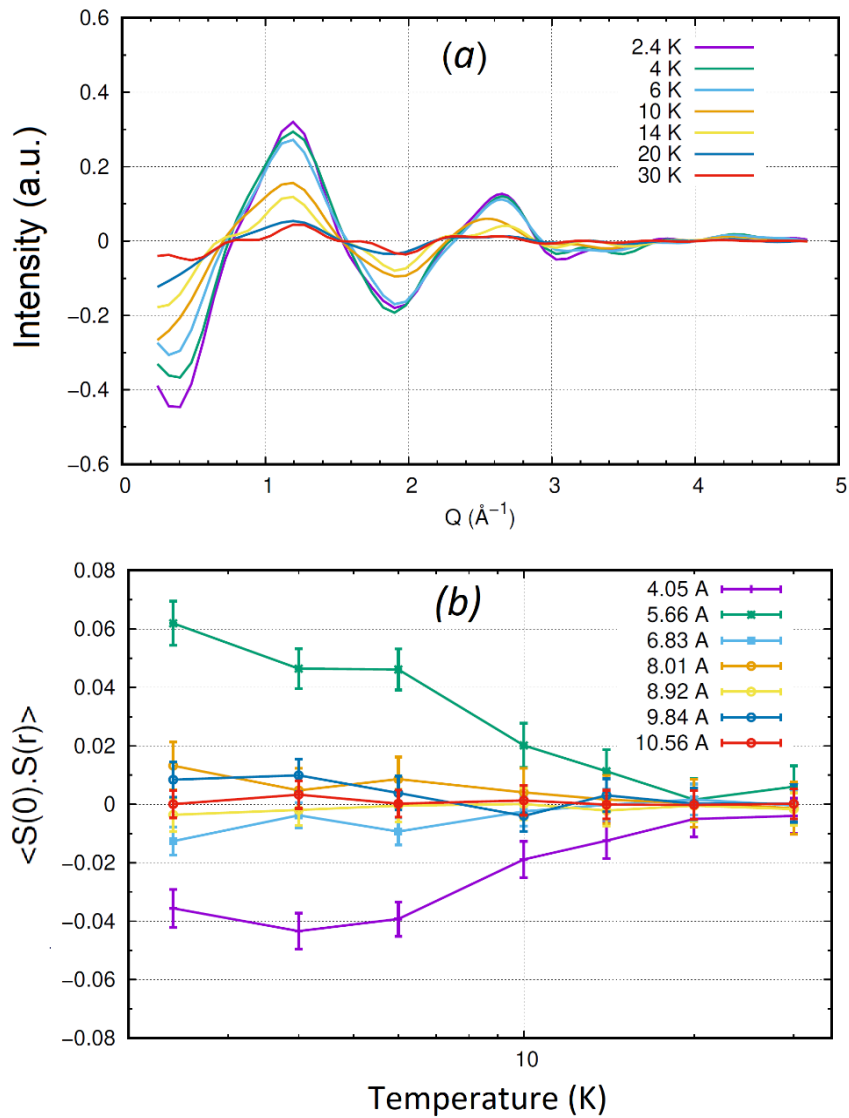

***Figure 7.*** Analysis of the magnetic diffuse scattering for EuFeAsF measured at D4. (*a*) Magnetic diffuse scattering obtained by subtracting the paramagnetic reference (diffraction pattern at 45 K) from the diffraction pattern at a given temperature. (*b*) Temperature dependence of the spin-correlation function for the given Eu-Eu distances.

**Microscopic analysis.**

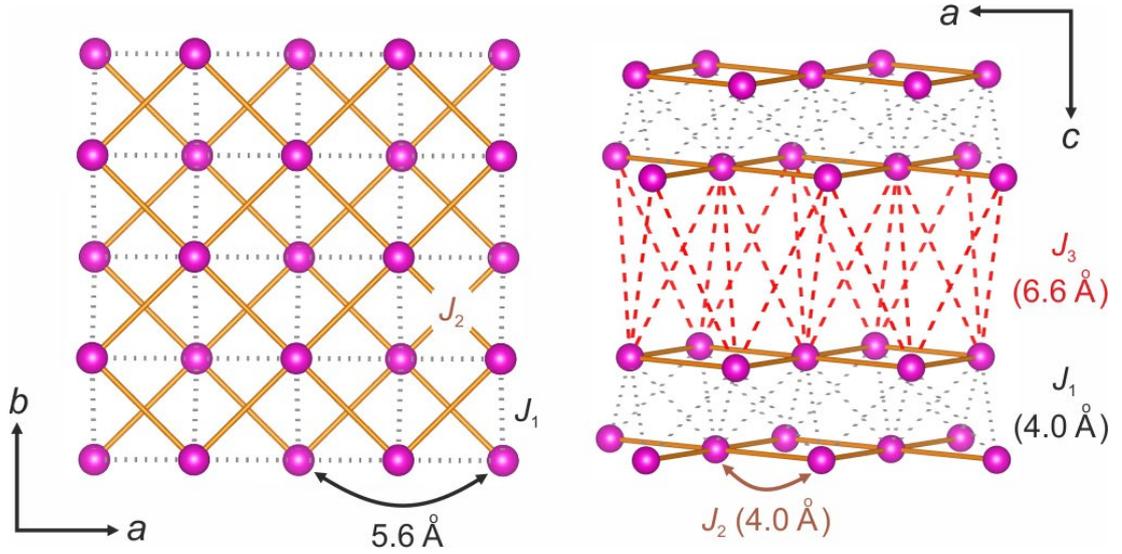

*Figure 8.* Schematic $Eu^{2+}$ spin lattice in Eu$T$AsF ($T$ = Zn, Mn, and Fe).

In Eu$T$AsF, the spin lattice of $Eu^{2+}$ features three main couplings: $J_1$, $J_2$, and $J_3$, as shown in *Figure 8*. We start by calculating these couplings in EuZnAsF where the [$T$As]$^-$ layer is nonmagnetic. Since the optimal DFT+$U$ parametrization for $Eu^{2+}$ is not known, we tested a wide range of the on-site Coulomb repulsion values ($U_{4f}$) while keeping the Hund's coupling fixed at $J_d = 1$ eV. The resulting magnetic couplings show a strong dependence on $U_{4f}$ (*Figure 9*), with $J_1$ dominant at $U_{4f} = 6$ eV and $J_2$ as well as $J_3$ dominant at $U_{4f} = 12$-$14$ eV. The experimentally relevant range of the $U_{4f}$ values can be chosen by calculating the saturation field for the (½ ½ 0) magnetic order of $Eu^{2+}$,

$$H_s = (g\, S\, /k_B)\, (2J_1 + 4J_2 + 2J_3) \qquad (1)$$

where $k_B$ is the Boltzmann constant, $S = 7/2$, and $g = 2$ is assumed because of the nearly quenched orbital moment of $Eu^{2+}$. A reasonable match with the experimental saturation field of about 4.5-5.0 T is reached for the $U_{4f}$ values of 10 eV or above. This value is somewhat higher than the screened on-site Coulomb repulsion of 6.5 - 7.5 eV reported for Gd[44] and Eu[45]. The discrepancy may be caused by the simplistic mean-field nature of DFT+U where only atomic 4f states are treated as correlated, and their mixing with other states is neglected. This comparison excludes the regime of dominant $J_1$ and suggests that both $J_2$ and $J_3$ should be the leading exchange couplings in EuZnAsF.

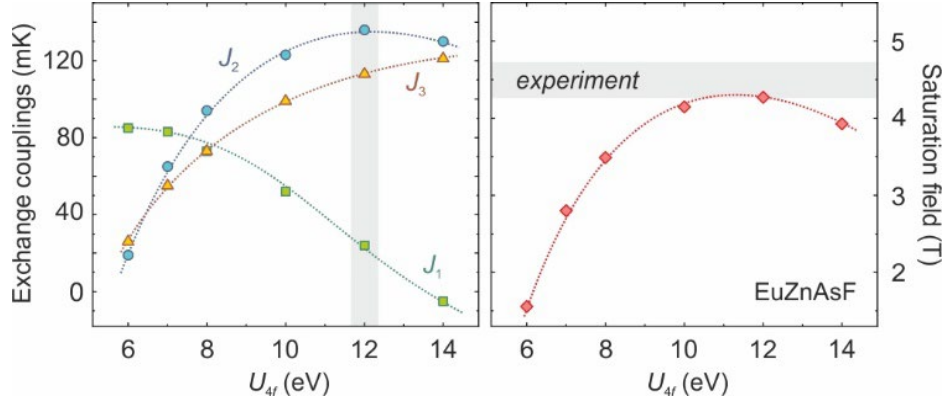

*Figure 9.* Left: exchange couplings in EuZnAsF as a function of $U_{4f}$, the Coulomb repulsion parameter of DFT+$U$. The gray bar indicates $U_{4f} = 12$ eV that was chosen for calculating exchange couplings in the Mn and Fe compounds. Right: saturation field estimated from Eq. (1). All lines are guides for the eye only.

For a further comparison with the experiment, we used experimental magnetic susceptibility and magnetization (*Figure 10*) that can be simultaneously fitted with the square-lattice model using $J_2$ = 160 mK and the g-value of 2.02. Together with the experimental (½ ½ 0) magnetic order, these fits indicate that $J_2$ should be the leading magnetic interaction in EuZnAsF and $U_{4f} > 10$ eV should be used for the evaluation of exchange couplings. Note that DFT+$U$ predicts a sizable $J_3$ in this regime, but attempts to introduce $J_3$ into the fits of thermodynamic properties led to a less favourable agreement with the experiment.

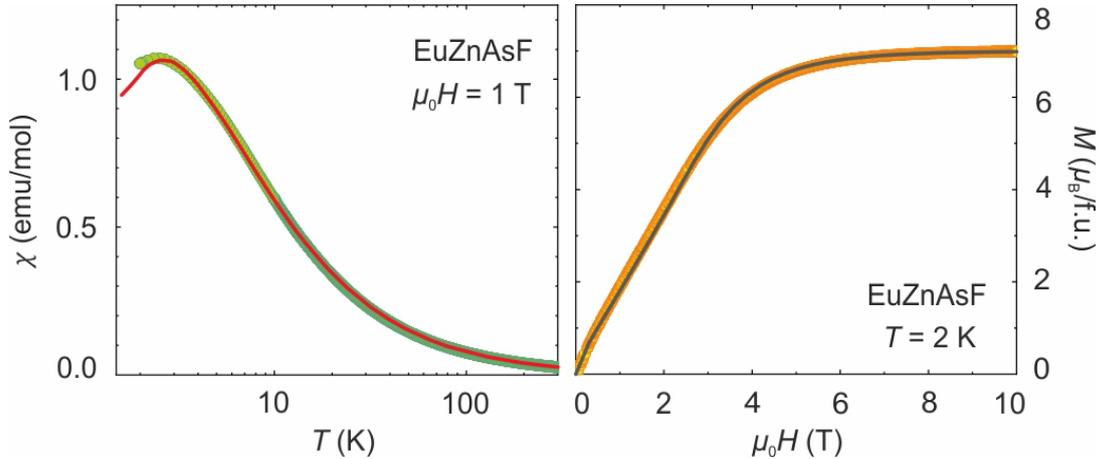

*Figure 10.* Thermodynamic properties of EuZnAsF (symbols) and the fits using the square-lattice model with $J_2 = 160$ mK ($g = 2.02$). Left: temperature-dependent magnetic susceptibility measured in the applied field of 1 T. Right: field-dependent magnetization measured at 2 K. The weak bend in $M(H)$ around 0.5 T is caused by an additional contribution that manifests itself by the 5 K anomaly in the low-field magnetic susceptibility (see text for details). This contribution was modelled with the Brillouin function.

In EuMnAsF and EuFeAsF, the [$T$As]⁻ layers become magnetic. The interactions between $3d$ ions are expected to be much stronger than those between Eu$^{2+}$. Therefore, the $T$-$T$ interactions become the dominant energy scale. In EuMnAsF, we evaluate them by placing Eu $4f$ states into the core, thus eliminating Eu$^{2+}$ magnetism. We find that leading Mn-Mn interactions are between nearest neighbours in the $ab$ plane, $J_1^{Mn} = 246$ K. Second-neighbour interactions are also significant, $J_2^{Mn}$

= 56 K, while the interplane interactions are very weak, $J_c^{Mn}$ = 0.1 K, and the local moment on Mn is about 3.8 $\mu_B$ in a good agreement with the experiment. This interaction regime is quite similar to the one reported earlier in other Mn-based 1111 and 112 compounds, [34–40,46] and results in the (0 0 ½) Néel order, in agreement with the experiment. In the case of EuFeAsF, a much smaller local moment of 1.5 $\mu_B$ was found on Fe atoms, and a full mapping of Fe-Fe interactions was not possible because this moment varied from one spin configuration to another. Nevertheless, we confirmed that stripe order with parallel spins along *a* and antiparallel spins along *b*, similar to SrFeAsF,[41] has the lowest energy.

*Table 2*. Eu-Eu exchange couplings in Eu*T*AsF calculated with $U_{4f}$ = 12 eV. The $J_i$ values are given in mK assuming $S$ = 7/2 for $Eu^{2+}$. In the case of EuFeAsF, the pairs of $J_1$ and $J_3$ values are due to the orthorhombic symmetry of the crystal structure.

|  | EuZnAsF | EuMnAsF | EuFeAsF |
| --- | --- | --- | --- |
| $J_1$ | 24 | 29 | 15/25 |
| $J_2$ | 136 | 82 | 170 |
| $J_3$ | 113 | 101 | 267/270 |

In addition to the *T-T* interactions, both Eu-*T* and Eu-Eu interactions are expected in EuMnAsF and EuFeAsF. However, both Néel (Mn) and stripe (Fe) orders in the 3*d* sublattice lead to a full cancellation of the Eu-*T* interactions because each $Eu^{2+}$ ion interacts with the Mn or Fe ions having the opposite spin. Therefore, on the level of Heisenberg interactions, the Eu and Mn (Fe) sublattices are fully decoupled. This decoupling is most tangible in the experimental magnetic structure of EuMnAsF where Eu and Mn spins are orthogonal to each other, and the two sets of magnetic reflections evolve independently.

We now evaluate Eu-Eu interactions in the *T* = Mn and Fe compounds using $U_{4f}$ = 12 eV as the optimal value for EuZnAsF. The resulting exchange couplings are summarized in *Table 2*. The overall $J_2$-$J_3$ scenario resembles the Zn compound. The value of $J_3$ is comparable in EuMnAsF and EuZnAsF and even increased especially in EuFeAsF. In all these cases, the interaction $J_3$ plays the same role as $J_1$ because it frustrates the square lattice, albeit by a link to the adjacent [EuF]$^+$ layer. It is then useful to introduce the frustration ratio $J_2/J_3$ that plays the same role as $J_2/J_1$ in the conventional frustrated-square-lattice model. The ratio decreases from 1.2 (EuZnAsF) to 0.8 (EuMnAsF) and 0.63 (EuFeAsF), thus approaching the value of 0.5 that separates different types of magnetic order in square-lattice antiferromagnets. Extensive theoretical studies show that collinear (½ ½ 0) phase is stable for $J_2/J_3$ > 0.60-0.65,[47,48] thus explaining the experimental $Eu^{2+}$ order in the Zn and Mn compounds. In EuFeAsF, the critical value of $J_2/J_3$ is approached, and incommensurate spiral states can become favourable, especially when they adapt to lattice distortions,[49] as present in EuFeAsF. Therefore, we conclude that the increasing $J_3$ and decreasing $J_2/J_3$ ratio should be the main causes for the incommensurate order of the $Eu^{2+}$ spins in EuFeAsF as opposed to the Mn and Zn compounds.

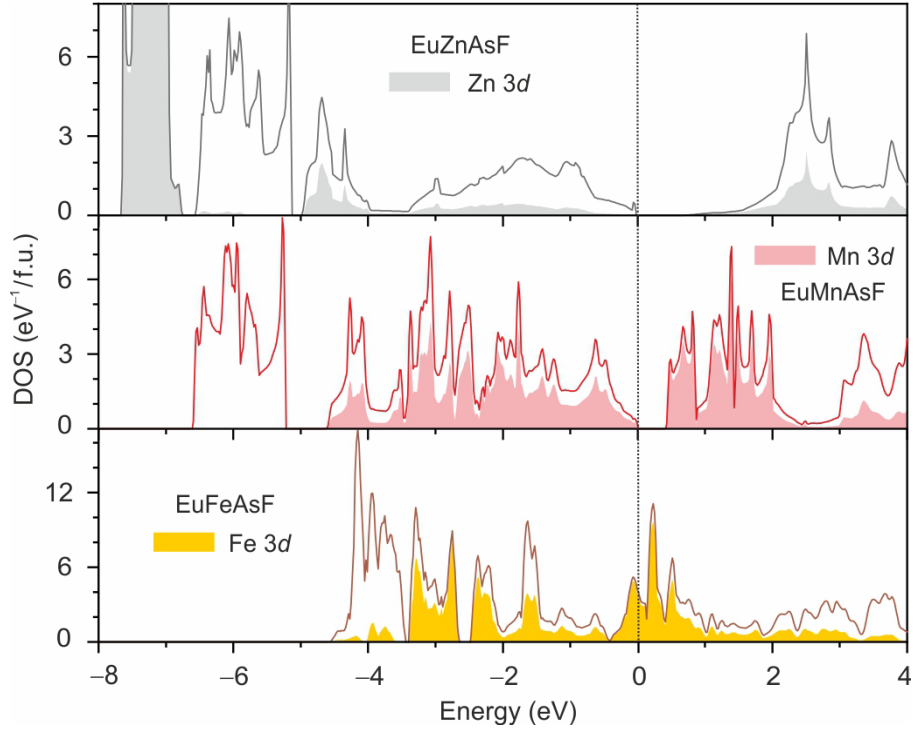

*Figure 11.* Electronic density of states for the Eu*T*AsF compounds. Colour filling shows the contribution of transition-metal 3*d* states. The Fermi level is at zero energy. Eu 4*f* states have been excluded, whereas relevant antiferromagnetic spin polarization of the transition-metal antifluorite layer has been introduced in the case of Mn and Fe.

The coupling regime of Eu*T*AsF contradicts general intuition, because short-range interactions $J_1$ are mostly inactive, while a significant interaction $J_3$ through the [*T*As]$^-$ slabs occurs. The size of this coupling seems to be controlled by the electronic structure of the [*T*As]$^-$ slab. In *Figure 11*, we compare electronic density of states for the three Eu*T*AsF compounds. EuZnAsF is semiconductor with the band gap of 0.6 eV. EuMnAsF features a smaller band gap of 0.4 eV and a much larger contribution of Mn 3*d* states near the Fermi level. Finally, EuFeAsF is metallic in agreement with the experimental resistivity data.[14,15] This metallicity is probably the main cause for the enhancement of $J_3$ and, thus, for the formation of the incommensurate Eu$^{2+}$ order in EuFeAsF.

**Conclusions**

Our neutron powder diffraction experiments reveal the nature of magnetically ordered phases in Eu*T*AsF (*T* = Zn, Mn and Fe) and explain the difference in their microscopic properties based on the DFT calculations. Whereas in semiconducting Eu*T*AsF (*T* = Zn and Mn), Eu$^{2+}$ magnetic moments form a commensurate ($k$ = (½ ½ 0)) stripe order pattern, the arrangement of magnetic moments in metallic EuFeAsF is incommensurate ($k$ = (0 0.961(1) ½)). Additionally, Mn$^{2+}$ (3$d^5$) magnetic moments in EuMnAsF are coupled by the strong antiferromagnetic interactions persisting at least up to room temperature, contrary to the case of Fe$^{2+}$ magnetism in EuFeAsF. We show that the $J_1$-$J_2$ magnetic interactions within the single [EuF] layer have comparable strength with the $J_3$ intralayer interaction in Eu*T*AsF (*T* = Zn and Mn). In EuFeAsF, the $J_3$ exchange is even enhanced due to the metallicity, which causes the deviations from the collinearity.

Supporting Information: magnetic properties and models of magnetic structures for EuFeAsF (DOC).

Acknowledgements

The neutron diffraction studies have been performed using beamtime provided by the ISIS and ILL facilities. We thank Dr. Navid Qureshi (ILL) for valuable discussion and development of the magnetic diffuse scattering option in MAG2POL software.